\documentclass[11pt]{article}

\usepackage[margin=1in]{geometry}
\usepackage{setspace}
\usepackage{graphicx}
\usepackage{booktabs}
\usepackage{longtable}
\usepackage{array}
\usepackage{amsmath}
\usepackage{natbib}
\usepackage{hyperref}

\usepackage{draftwatermark}

\SetWatermarkText{arXiv Version by Kaushik Dutta, USF}
\SetWatermarkScale{0.8}
\SetWatermarkColor[gray]{0.8}

\graphicspath{{figures/}}
\title{From Precision Medicine to Precision Education:\\
A Vision for AI-Powered Student Digital Twins, Preventive Student Success, and Career-Aligned Academic Pathways}

\author{ Kaushik Dutta\\ University of South Florida, Tampa, Florida, USA\\ \texttt{duttak@usf.edu} }
\date{}

\begin{document}

\maketitle

\begin{abstract}
Higher education remains largely reactive in its approach to student success. Institutions frequently identify academic problems only after students have failed courses, fallen behind in degree progression, accumulated excessive debt, or departed without a credential. Healthcare faced a similar challenge decades ago. It responded by shifting from reactive treatment to preventive care powered by predictive models, risk stratification, electronic health records, and artificial intelligence (AI). This paper argues that higher education stands at an analogous inflection point. Drawing on advances in learning analytics, educational data mining, machine learning, workforce analytics, and digital twin technologies, we propose a paradigm we call \textit{Precision Education}. Under this framework, AI continuously analyzes academic, behavioral, financial, and career data to identify emerging risks, recommend personalized interventions, optimize educational pathways, and align academic decisions with long-term career success. Central to the model is the Student Digital Twin, a continuously updated representation of a learner that can simulate multiple educational futures and intervention scenarios. We ground the vision in evidence from early deployments such as Course Signals at Purdue and GPS Advising at Georgia State University. We also argue that prediction alone is insufficient. The central methodological challenge is the move from prediction to causal, actionable intervention. The paper presents a conceptual framework, examines enabling technologies, reviews the empirical record and its limits, analyzes ethical and governance implications, and outlines a research agenda for the next decade of AI-enabled higher education.
\end{abstract}


\section{Introduction}

Modern healthcare has changed its philosophy of care. Rather than waiting for catastrophic outcomes, physicians use data to identify risk factors and intervene before serious problems emerge. Cardiovascular disease, diabetes, cancer, and mental health disorders are increasingly managed through prediction and prevention rather than late-stage treatment. Precision medicine extends this logic further. It targets interventions to the individual rather than the population average.

Higher education has not undergone a comparable transformation. Most universities still manage student success through retrospective reporting. Metrics such as retention rates, graduation rates, progression ratios, DFW rates, and student debt loads are typically measured after outcomes have already occurred. As a result, institutional interventions often arrive too late to change a student's trajectory. Purdue University found this directly. Its paper-based faculty warnings often came too late to be effective, and the messages sent to students were too general to prompt behavior change \citep{arnold2012course}.

The emergence of AI, learning analytics, and educational data mining creates an opportunity to reimagine this model. Research over the past fifteen years shows that machine learning can predict student attrition, academic risk, and course performance with useful accuracy \citep{berens2019early,beaulac2019predicting}. Existing research and articles document steady advances in dropout prediction, early-warning systems, and outcome forecasting.

Yet prediction alone does not create value. A risk score changes nothing on its own. The value lies in an educational ecosystem that can identify risk before failure occurs, select an intervention that actually improves outcomes, and align academic decisions with a student's long-term goals. The distance between an accurate prediction and an effective action is the core problem this paper addresses.

\begin{figure}[htbp]
\centering
\includegraphics[width=0.95\textwidth]{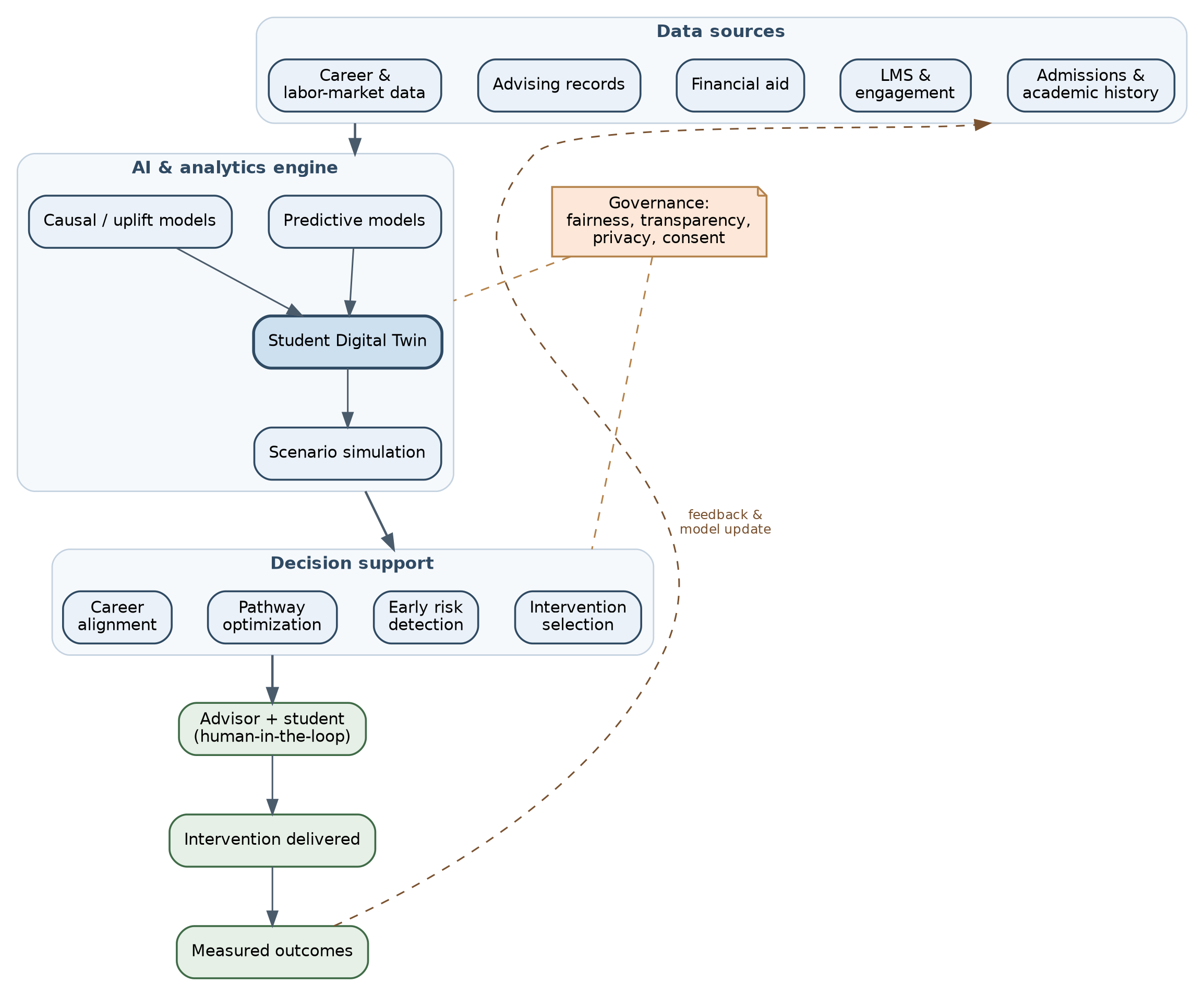}
\caption{The Precision Education framework}
\label{fig:framework}
\end{figure}

We argue that higher education should adopt a healthcare-inspired paradigm of preventive student success, which we call Precision Education. The framing is not only ours. Timothy Renick, who led Georgia State University's student success reforms, describes the model in explicitly medical terms. He argues that the earlier a problem is identified, the more likely a solution becomes, and that unattended problems worsen over time. This paper formalizes that intuition, grounds it in the empirical record, and extends it toward a future of continuously updated Student Digital Twins. Figure~\ref{fig:framework} gives the overall framework.

\section{From Learning Analytics to Precision Education}

Learning analytics emerged as a field focused on collecting and analyzing educational data to improve teaching and learning \citep{siemens2011penetrating}. Educational data mining developed complementary machine learning methods to identify patterns in behavior and performance \citep{baker2014educational, Romero_2020}. Early proofs of concept showed that data from a learning management system could power an early-warning system for instructors \citep{macfadyen2010mining}.

More recently, researchers have advanced the concept of Precision Education. Stephen Yang introduced the term in a keynote at the International Conference on Computers in Education, framing it as a new challenge for AI, machine learning, and learning analytics \citep{yang2019precision}. Yang and colleagues argue that the goal of Precision Education is not merely to understand student behavior. The goal is to identify at-risk students as early as possible and provide timely, tailored intervention \citep{yang2021precision,luan2021review,tsai2020precision}. The framing deliberately pairs data science with what Yang calls warm humanity. Technology identifies the signal. Human judgment shapes the response.

The evolution can be organized across five stages. The first four are widely used in analytics maturity models. We add a fifth to capture the autonomous, continuously operating future state.

\begin{enumerate}
\item[\textbf{Stage 1:}] \textbf{Descriptive analytics.} What happened? How many students graduated? Which courses have the highest failure rates?
\item[\textbf{Stage 2:}] \textbf{Diagnostic analytics.} Why did it happen? Which factors contributed to failure or persistence?
\item[\textbf{Stage 3:}] \textbf{Predictive analytics.} What is likely to happen next? Which students are at risk? Most machine learning research in this area sits here. Studies show that models using academic, demographic, and behavioral features can flag students at risk of attrition \citep{berens2019early}.
\item[\textbf{Stage 4:}] \textbf{Prescriptive and preventive analytics.} What should we do? Which intervention is most likely to improve outcomes? What pathway is optimal for this student? This stage requires causal reasoning, not only correlation.
\item[\textbf{Stage 5:}] \textbf{Autonomous and adaptive student success.} Can the system monitor continuously, simulate alternatives, recommend, learn from outcomes, and update itself, with humans in the loop for consequential decisions? This is the frontier that digital twins and agentic AI make plausible.
\end{enumerate}

\begin{figure}[htbp]
\centering
\includegraphics[width=0.95\textwidth]{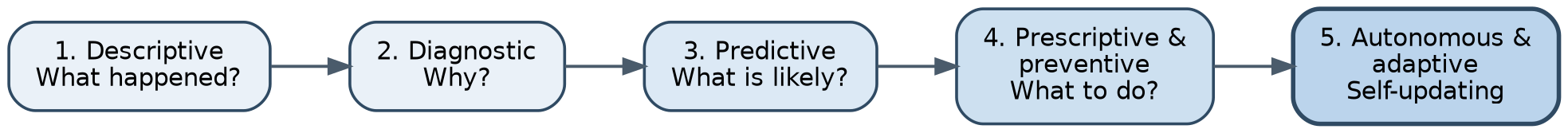}
\caption{Analytics maturity progression}
\label{fig:maturity}
\end{figure}

The field has invested heavily in Stage 3. The value proposition of Precision Education lives in Stages 4 and 5.

\section{A Healthcare Model for Student Success}

The analogy between healthcare and higher education is instructive because the workflow maps closely.

\begin{table}[htbp]
\centering
\caption{Healthcare and higher education workflow analogy}
\label{tab:healthcare}
\begin{tabular}{ll}
\toprule
\textbf{Healthcare} & \textbf{Higher Education} \\
\midrule
Risk screening & Academic risk detection \\
Disease prediction & Attrition prediction \\
Diagnosis & Root-cause analysis of struggle \\
Treatment planning & Intervention planning \\
Care pathway optimization & Degree pathway optimization \\
Preventive medicine & Preventive student success \\
Patient monitoring & Student monitoring \\
Clinical decision support & Advising decision support \\
Comparative effectiveness research & Intervention effectiveness research \\
\bottomrule
\end{tabular}
\end{table}

In healthcare, clinicians combine patient history, diagnostic testing, behavioral data, and predictive models to estimate future outcomes and to choose treatments. Universities hold remarkably similar datasets. These include admissions records, academic histories, learning management system interactions, financial aid information, registration patterns, advising notes, engagement activity, and employment outcomes.

The difference is not the availability of data. The difference is how institutions use it. Healthcare increasingly uses data to change future outcomes. Higher education still primarily uses data to explain past outcomes.

The analogy also carries warnings. Healthcare learned that prediction without a validated treatment protocol produces anxiety, not health. It learned that models trained on one population can perform poorly on another. It learned that screening tools can encode and amplify bias. Precision Education must import these lessons along with the ambition.

\section{Evidence from Early Deployments}

The vision is not speculative. Two large deployments show what preventive student success can look like at scale, and both reveal the limits of the current state of the art.

\subsection{Course Signals at Purdue}

Purdue piloted Course Signals in 2007 and reported it as one of the first operational learning analytics systems \citep{arnold2012course}. The system combined grades, prior academic history, demographic characteristics, and effort measured through learning management system activity. It produced a red, yellow, or green signal on the student's course page and prompted personalized faculty outreach. Purdue reported gains in help-seeking, lower D and F rates, and improved retention. These reported outcomes were later questioned on methodological grounds. Critics argued that some retention gains reflected the fact that students who took more Course Signals courses were, by construction, students who persisted longer, a selection effect rather than a treatment effect \citep{caulfield2013signals}. The episode is a cautionary anchor for this paper. An impressive correlation between an intervention and an outcome is not evidence that the intervention caused the outcome.

\subsection{GPS Advising at Georgia State University}

Georgia State launched GPS Advising in 2012. The system tracks more than 40,000 undergraduates daily against roughly 800 risk factors and sends advisors alerts when a student goes off path \citep{georgiastate_gps, Page2017AICollege}. The institution reports more than 250,000 advisor meetings prompted by alerts, with tens of thousands each year \citep{georgiastate_gps, Rossman2022MicroGrants}. Georgia State pairs prediction with two other elements that matter as much as the model. First, it added advising capacity so that alerts led to real human contact within about 48 hours. Second, it added Panther Retention Grants, small completion grants for students who were academically on track but held a modest unpaid balance. Georgia State reports that six-year graduation rates rose over the period and that graduation gaps by race, ethnicity, and income narrowed sharply, in some reports to near parity. The university also deployed an AI chatbot, Pounce, to reduce summer melt and answer enrollment questions at scale.

Two lessons follow. First, the model is the smallest part of the system. Organizational design, advising capacity, and financial supports carried much of the effect. Second, the strongest reported outcome of these deployments is not attrition prediction accuracy. It is the narrowing of equity gaps, which is precisely the outcome most at risk from biased models. That tension motivates Sections~\ref{sec:career} and \ref{sec:agentic}.

Beyond these flagship cases, a broad research literature demonstrates predictive feasibility. Administrative-data models identify likely dropouts at German universities early in enrollment \citep{berens2019early}. Random forests predict not only whether a student will succeed but also which major fits their profile \citep{beaulac2019predicting}. This second result is important. It hints at a question richer than at-risk flagging, which we develop in Section~\ref{sec:programs}.

\section{The Student Digital Twin}

The most transformative concept emerging from the convergence of AI and education is the Student Digital Twin.

A digital twin is a virtual representation of a real-world system that is continuously updated through data streams and used to simulate future states and optimize performance. Digital twins are established in manufacturing, aerospace, healthcare, and urban planning. Research has begun to extend the concept to education, both for personalized learning environments and for lifelong learning \citep{abdelaziz2024preprint,kinsner2021digital,almawadieh2024digital,Fuller2020DigitalTwin}.

We extend the concept to student success. A Student Digital Twin is a continuously evolving model of an individual learner's academic performance, learning behavior, career interests, degree progression, financial circumstances, co-curricular engagement, and workforce preparedness. Its purpose is not to display a dashboard. Its purpose is to simulate.

The twin should support counterfactual questions:

\begin{itemize}
\item What happens if the student stays on the current pathway?
\item What happens if a gateway course is repeated with tutoring support?
\item What happens if academic coaching is introduced now versus next term?
\item What happens if the student changes majors, and to which major?
\item What happens if internship participation increases?
\end{itemize}

\begin{figure}[htbp]
\centering
\includegraphics[width=0.78\textwidth]{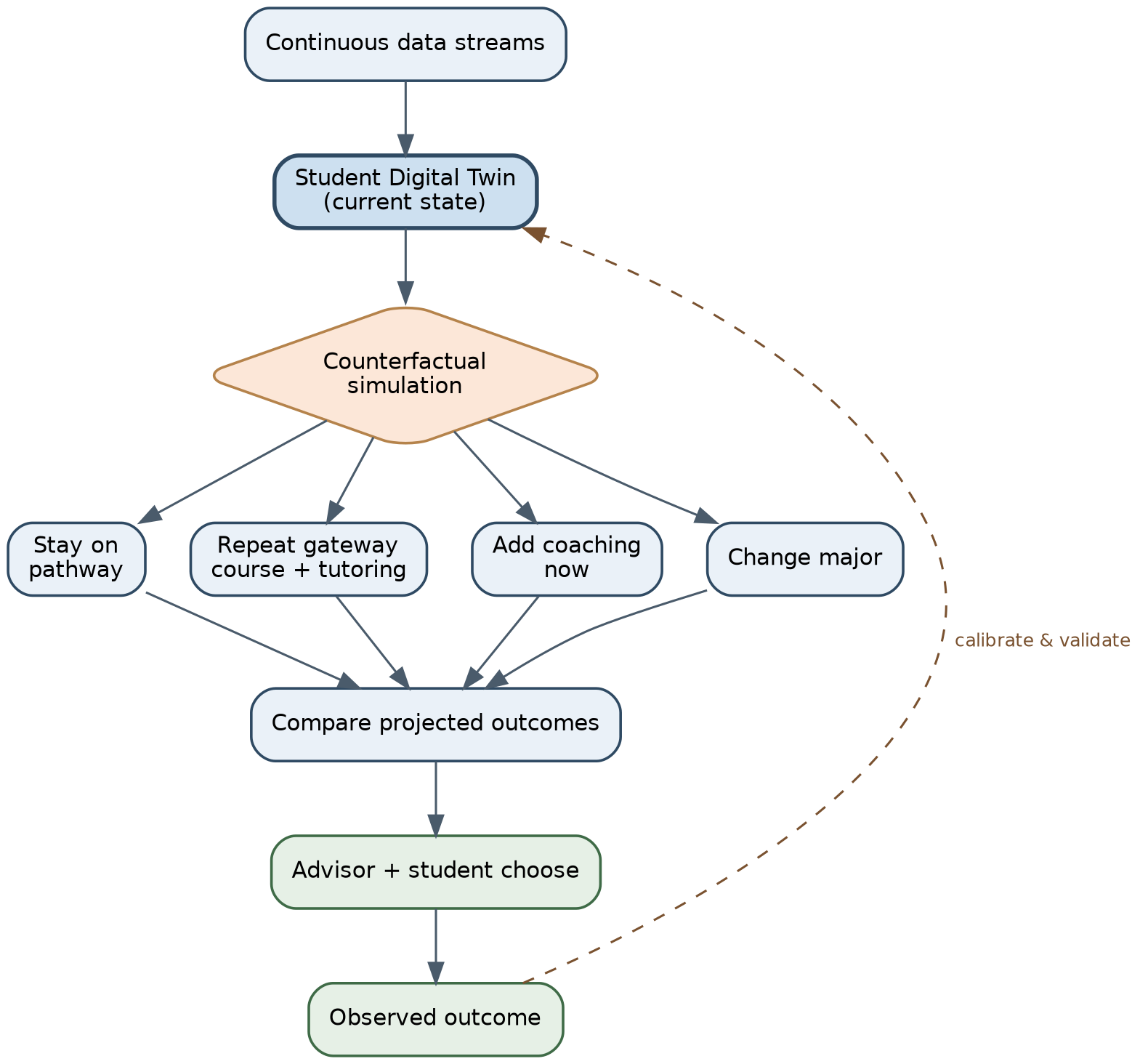}
\caption{Student Digital Twin}
\label{fig:twin}
\end{figure}

Figure~\ref{fig:twin} shows the twin as a simulation engine rather than a dashboard.

This reframes advising. Advising moves from retrospective consultation toward future-state simulation. The advisor and student examine several plausible futures together and choose among them.

\subsection{Fidelity, Calibration, and Validation}

A twin is only as useful as it is trustworthy. Three design properties deserve attention.

First, fidelity should be matched to purpose. A low-fidelity twin that models term-to-term progression may be sufficient for advising triage. A high-fidelity twin that models course-level mastery requires far more data and computation. Recent work proposes multi-fidelity designs that align model complexity to the decision at hand, and even to learning frameworks such as Bloom's taxonomy \citep{cao2021}. Institutions should not build a high-fidelity twin where a low-fidelity one would do.

Second, the twin must be calibrated. A predicted 70 percent probability of passing should correspond to roughly 70 percent observed pass rates. Calibration is distinct from accuracy and is frequently neglected. Advisors will lose trust in a twin whose probabilities do not mean what they say.

Third, simulations must be validated against real outcomes over time. A twin that predicts the effect of an intervention makes a causal claim. That claim needs empirical support, ideally from experiments or strong quasi-experiments, not only from historical correlations. Without this, the twin projects confident futures that were never tested.

\subsection{The Sim-to-Real Gap and Conceptual Caution}

Digital twin language is seductive, and the education literature is still young and partly aspirational. Some scholars warn that the discourse conflates personalization, which centers the learner, with personalization delivered by systems, which can center the vendor and the platform \citep{arantes2024digital, ZawackiRichter2019AIHE, luan2021review}. A Student Digital Twin models a person. It is not the person. It omits context, motivation, relationships, and circumstance that no data stream captures. The model should be treated as a decision aid that expands human judgment, not as an oracle that replaces it.

\section{From Prediction to Causal Intervention}

This section states the paper's central methodological argument. Prediction and intervention are different problems, and higher education has confused them.

Predictive models answer a correlational question. Given this student's features, how likely is a given outcome? Intervention requires a causal question. If we act, how much will the outcome change, and for whom? A model can predict attrition with high accuracy and still provide no guidance about what to do, because the features that predict best, such as prior GPA, are often the features an institution cannot change.

\begin{figure}[htbp]
\centering
\includegraphics[width=0.82\textwidth]{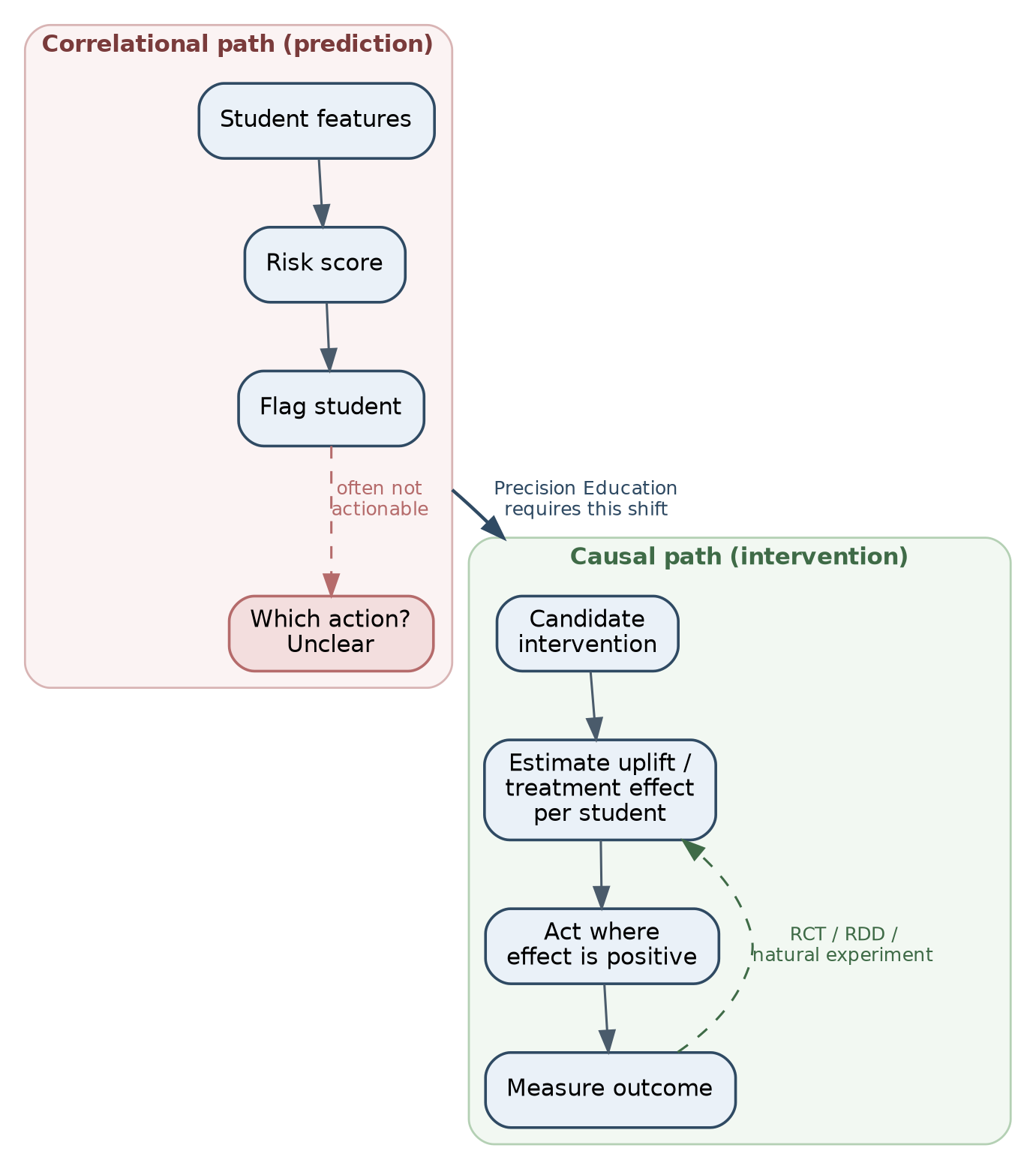}
\caption{Prediction versus causal intervention}
\label{fig:causal}
\end{figure}

Precision medicine faced the identical problem and answered it with comparative effectiveness research, randomized trials, and causal inference. Precision Education needs the same toolkit. Figure~\ref{fig:causal} contrasts the two paths.

Several methods apply. Uplift modeling, also called heterogeneous treatment effect estimation, estimates the incremental effect of an intervention on each individual rather than the average effect on the group. It distinguishes students who will benefit from coaching from students who would have persisted anyway and from students for whom coaching does nothing. Randomized encouragement designs and staggered rollouts let institutions estimate causal effects while still serving students. Quasi-experimental methods such as regression discontinuity fit naturally where thresholds already exist, for example probation cutoffs or grant eligibility. The Georgia State Panther Retention Grants program, which triggers at a balance threshold, is a natural setting for such designs.

The Student Digital Twin is best understood as a structured way to reason about counterfactuals. Each simulated future is a causal claim about an intervention. The scientific credibility of the twin therefore depends on the quality of the causal evidence behind its edges, not on the sophistication of its interface.

Two failure modes make this argument urgent. The first is the self-fulfilling prophecy. If a low prediction lowers the support a student receives, the prediction can cause the outcome it forecast. The second is Goodhart's law. When a predictive metric becomes a target, actors optimize the metric rather than the underlying goal, and the metric degrades. Both failure modes are invisible to accuracy metrics and only become visible when institutions ask the causal question directly.

\section{Predicting Success Across Programs, Not Only Within One}
\label{sec:programs}

Current student success systems usually ask a single question. Will this student succeed in this program? A more valuable question is often different. Where is this student most likely to succeed?

The distinction is consequential. Consider a student struggling in engineering. Traditional early-warning systems focus on interventions that keep the student in engineering. A Student Digital Twin can evaluate historical performance patterns, strengths and weaknesses, peer-outcome comparisons, career aspirations, and alternative pathways. Empirically, models already predict best-fit majors from student records, not only pass or fail within a chosen major \citep{beaulac2019predicting, Bettinger2014StudentCoaching}. The system might discover that students with similar profiles reached comparable or better career outcomes through data analytics, information systems, business analytics, or health informatics.

This capability is powerful and dangerous in equal measure, so its ethics must be stated up front. The goal is not to steer students away from challenging or high-status fields. Doing so along demographic lines would reproduce exactly the tracking harms that Section~\ref{sec:ethics} warns against. The goal is to expand the set of futures a student can see and choose among. Recommendations should widen options, not narrow them, and the student, not the model, should decide. A twin that quietly nudges first-generation or minoritized students out of engineering is not Precision Education. It is algorithmic tracking with a friendlier interface.

\section{Career-Centric Academic Planning}
\label{sec:career}

Traditional educational planning begins with a program. Students choose a major and then explore careers. AI enables the reverse. A student can begin with a career aspiration, and the system can work backward to the courses, experiences, and credentials that reach it.

This reversal depends on infrastructure that did not exist a decade ago. Occupational and skills taxonomies now make it possible to connect learning to work at fine granularity. The United States Department of Labor's O*NET-SOC taxonomy classifies roughly one thousand occupations by their knowledge, skills, abilities, and tasks \citep{onet2024taxonomy}. In Europe, the ESCO framework \citep{EuropeanCommissionESCO} links occupations, skills, and qualifications across member states. Commercial providers add real-time granularity. Lightcast maintains an open skills taxonomy of more than thirty thousand skills \citep{lightcast2025skills}. These taxonomies give every occupation a machine-readable skill profile that can be aligned against a curriculum.

\begin{figure}[htbp]
\centering
\includegraphics[width=0.95\textwidth]{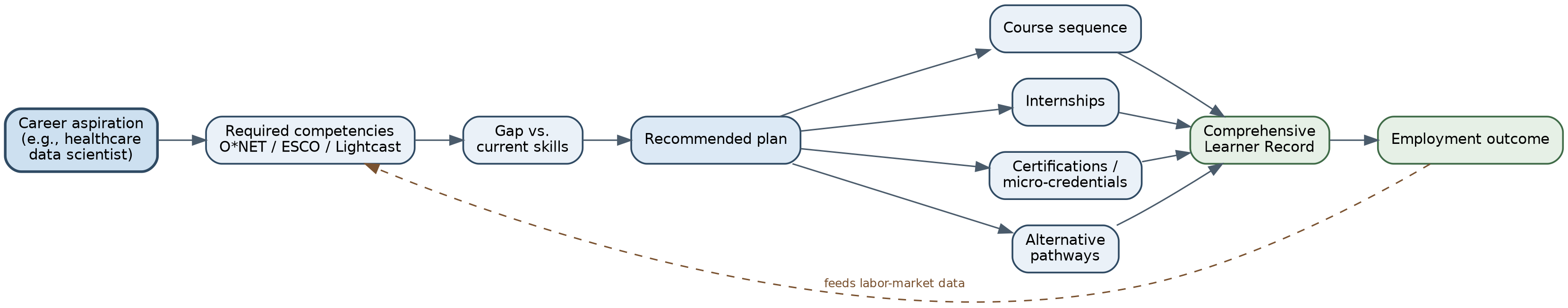}
\caption{Career-centric, or reverse, academic planning}
\label{fig:career}
\end{figure}

Large language models add a further layer. They can parse job descriptions, map them to skills, compare those skills against course learning outcomes, and surface gaps. A student who says, ``I want to become a healthcare data scientist,'' can receive a concrete map. The system identifies required competencies, relevant programs, an efficient course sequence, recommended internships, useful certifications, and alternative pathways to the same destination. Figure~\ref{fig:career} shows this reverse planning pipeline.

Several extensions follow.

\paragraph{Stackable and micro-credentials.} Skills maps naturally support stackable credentials and micro-credentials that let students accumulate labeled competencies rather than only a single terminal degree. This matters for working and returning learners.

\paragraph{The Comprehensive Learner Record.} A skills-based transcript, sometimes called a Comprehensive Learner Record, can carry verified competencies alongside courses, giving employers a richer signal and giving students a portable record of what they can do.

\paragraph{Return on education.} Labor market data allow estimates of expected earnings trajectories by pathway. Combined with cost and debt data, this supports honest conversations about return on education. It also introduces risk. Earnings-maximizing recommendations can devalue fields that are socially important but modestly paid. The system should inform choices about value, not dictate them by salary alone.

The development of new AI and analytics programs has already incorporated workforce demand, labor market intelligence, and skills analysis into program design. This experience suggests that career-aligned planning is feasible at the program level today and can be pushed toward the individual level as twins mature.

\section{Autonomous and Agentic Student Success Systems}
\label{sec:agentic}

The future state may involve continuously operating platforms that combine several capabilities. Early risk detection identifies challenges before failure. Personalized intervention selection matches students to actions that are effective for people like them. Career pathway optimization aligns academic choices with workforce opportunity. Academic journey simulation models multiple futures. Dynamic advising updates recommendations as new data arrive.

Generative AI and agentic systems make an interactive advising layer plausible. Early studies explore ChatGPT and similar models as advising aids and find promise for routine guidance alongside clear limits for complex, emotional, or high-stakes conversations \citep{akiba2023advising, Page2017AICollege}. Georgia State's Pounce chatbot demonstrated measurable value for well-scoped tasks such as reducing summer melt.

Three design constraints matter. First, retrieval-augmented generation should ground the advising agent in the institution's own catalog, policies, and the student's record, which reduces hallucination and keeps advice actionable. Second, the agent should escalate to a human advisor for consequential or sensitive decisions rather than resolving them autonomously. Third, the agent's recommendations should be traceable to sources and to model logic, so that students and advisors can see why a recommendation was made. The convergence of institutional data, workforce analytics, generative AI, and predictive modeling makes these systems realistic, but only if they are built to augment advisors rather than to replace the human relationship that carried much of the effect at Georgia State.

\section{Ethical, Governance, and Policy Challenges}
\label{sec:ethics}

The same technologies that enable transformative student success also create serious risks. Reviews of AI in education emphasize concerns about algorithmic bias, explainability, transparency, fairness, privacy, and accountability \citep{baker2022algorithmic,kizilcec2022fairness}. We organize these into principles and then name the specific failure modes that Precision Education must guard against.

\subsection{Governing Principles}

\begin{enumerate}
\item \textbf{Human in the loop.} AI should support advisors, not replace them. Consequential decisions require human judgment.
\item \textbf{Transparency and explainability.} Students should be able to understand how a recommendation was generated. Explainability is not a courtesy. In advising it is a condition of trust and of contestability. Methods that make model reasoning legible to non-experts are a research priority in their own right.
\item \textbf{Student agency.} Recommendations should expand choices, not restrict them. The student decides.
\item \textbf{Fairness auditing.} Models must be monitored continuously for disparate performance and disparate impact across groups, not audited once at launch.
\item \textbf{Data minimization and consent.} Only necessary data should be used, under clear consent, consistent with FERPA in the United States and comparable regimes elsewhere. Student data collected for support should not silently become data used for surveillance or enforcement.
\item \textbf{Outcome accountability.} Institutions must evaluate whether interventions actually improve outcomes, using causal methods, and must retire interventions that do not.
\end{enumerate}

\subsection{Specific Failure Modes}

\paragraph{Bias in at-risk models.} Baker and Hawn review concrete evidence that educational algorithms perform unequally across race, ethnicity, gender, nationality, and disability, and note that many biases remain unknown because they have not yet been studied \citep{baker2022algorithmic}. A common cause is convenience sampling, where models trained on easily available data perform worst for underrepresented groups. Because the strongest documented benefit of predictive advising is narrowing equity gaps, a biased model does not merely fail. It inverts the goal.

\paragraph{Algorithmic tracking.} Cross-program recommendation can degenerate into steering students out of fields along demographic lines. This is the historical harm of tracking, rebuilt in software. Fairness auditing must specifically test whether recommendations differ by protected group after controlling for legitimate factors.

\paragraph{Self-fulfilling prophecy and labeling.} A risk label can change how faculty and advisors treat a student and can lower the support offered, causing the predicted outcome. Systems should surface support opportunities, not deficit labels, and should measure whether being flagged helps or harms.

\paragraph{Opacity and contestability.} Course Signals used a proprietary, unpublished algorithm \citep{arnold2012course, Bettinger2014StudentCoaching}. Opaque high-stakes models are hard to audit and hard to contest. Transparency should be a procurement requirement, not an afterthought.

\paragraph{Equity of access to the technology itself.} Well-resourced institutions can build twins and advising platforms. Under-resourced institutions, which enroll many of the students who would benefit most, may not. Precision Education could widen institutional inequality even as it narrows within-institution gaps. Shared infrastructure, open models, and consortium approaches are one response.

The future of educational AI will depend as much on governance as on technology.

\section{Institutional and Organizational Readiness}

Technology is necessary but far from sufficient. The Georgia State case shows that the model was the smallest component. The effect came from added advising staff, defined response protocols, financial supports, and sustained leadership. Precision Education is an organizational transformation, not a software purchase.

Several readiness conditions follow. Data infrastructure must integrate admissions, student information, learning management, financial aid, advising, and outcomes systems into a governed, timely, and reliable pipeline. Advising capacity must be sufficient to respond to what the system detects, since an alert with no responder is worse than no alert. Institutional research and data governance functions must own model validation, fairness auditing, and outcome evaluation. Change management must bring faculty and advisors into the design, because a system that advisors do not trust will not be used. An analytics maturity progression, moving deliberately from descriptive reporting toward validated prescriptive action, gives institutions a realistic sequence rather than an all-at-once leap.

Implementation science offers useful discipline here. It treats the rollout of an intervention as itself a research object, with attention to fidelity, adaptation, and sustainability. Precision Education programs should evaluate not only whether a model is accurate but whether the surrounding human system delivers the intervention as intended.

\section{Research Agenda}

Future research should address several foundational questions. We state them with methodological specificity so they are actionable.

\begin{itemize}
\item \textbf{Digital twin design and validation.} How should Student Digital Twins be architected, and how should simulated counterfactuals be validated against experimental or quasi-experimental outcomes rather than historical correlations?
\item \textbf{Causal intervention effectiveness.} Which interventions produce measurable causal improvements, for whom, and at what cost? Uplift modeling, randomized encouragement designs, and regression discontinuity at existing thresholds are natural methods.
\item \textbf{Career integration.} How should labor market outcomes and skills taxonomies be integrated into educational optimization without reducing education to earnings maximization?
\item \textbf{Fairness and governance.} What continuous auditing protocols, transparency requirements, and contestability mechanisms ensure that predictive advising narrows rather than widens equity gaps?
\item \textbf{Explainability for non-experts.} How can model reasoning be made legible and contestable for students and advisors, not only for data scientists?
\item \textbf{Privacy-preserving analytics.} How can institutions gain predictive power while protecting students, using consent, data minimization, and techniques such as federated learning across institutions?
\item \textbf{Organizational operationalization.} What structures, staffing, and change processes are required to run Precision Education at scale, and how do effects depend on them rather than on the model?
\item \textbf{Access equity.} How can under-resourced institutions gain these capabilities through shared or open infrastructure so the technology narrows rather than widens gaps between institutions?
\end{itemize}

These questions span Information Systems, Learning Analytics, Machine Learning, Higher Education, and Public Policy, and they invite exactly the kind of interdisciplinary work that AI fairness and explainability research is positioned to lead.

\section{Tensions and Limitations}

Intellectual honesty requires naming the tensions this vision does not resolve.

Prediction accuracy and actionability pull in different directions. The most predictive features are often the least changeable. Personalization and privacy trade off. Richer twins require more data and more exposure. Efficiency and equity can conflict. Optimizing aggregate graduation rates can disadvantage the students who need the most support. Automation and relationship can conflict. The human advising contact that carried much of Georgia State's effect is expensive and does not scale as cheaply as a chatbot. Finally, the evidence base remains thin. Much of the flagship data is institution-reported and correlational, and independent causal evaluation is limited. This paper proposes a direction and a research program, not a finished proof.

\section{Conclusion}

Healthcare transformed itself by moving from treatment to prevention, and precision medicine went further by treating individuals rather than averages. Higher education now faces a similar opportunity and a similar set of hazards.

Artificial intelligence, learning analytics, workforce intelligence, and digital twin technologies make it possible to build individualized educational pathways that adapt continuously to student needs, circumstances, and aspirations. The vision extends beyond retention dashboards and predictive flags. It proposes a future in which every student has a continuously evolving digital representation that can identify risk, simulate alternative futures, recommend interventions grounded in causal evidence, and align academic choices with personal and professional goals.

The lesson of the early deployments is that the technology is the smaller half of the work. The larger half is the human system around it, and the discipline to ask not only what will happen but whether our actions actually change it, and for whom. Precision medicine transformed healthcare by treating individuals rather than averages, and by demanding evidence that treatments work. Precision Education can transform higher education on the same terms, guiding each student along a uniquely personalized and continuously validated path toward graduation, career success, and lifelong learning.

\bibliographystyle{apalike}
\bibliography{references}

\end{document}